\documentclass[twocolumn,prl,preprintnumbers,showpacs,bibnotes]{revtex4}

\usepackage[latin1]{inputenc}
\usepackage{amsmath,amssymb,bbm,epsfig}

\newcommand{\assign}{:=}
\newcommand{\bigintlim}{\int}
\newcommand{\bignone}{\,}
\newcommand{\op}[1]{#1}
\newcommand{\mathpi}{\pi}
\newcommand{\mathd}{\mathrm{d}}
\newcommand{\mathe}{\mathrm{e}}
\newcommand{\tmem}[1]{{\em #1\/}}
\newcommand{\tmmathbf}[1]{\ensuremath{\boldsymbol{#1}}}
\newcommand{\tmop}[1]{\ensuremath{\operatorname{#1}}}

\newcommand{\emdash}{---}

\begin{document}
\author{Klaus Hornberger}
\affiliation{Arnold Sommerfeld Center for Theoretical Physics,
Ludwig-Maximilians-Universität München, Theresienstraße 37, 80333 Munich, Germany}
\homepage{www.klaus-hornberger.de}
\preprint{Phys. Rev. Lett. \textbf{97}, 060601 (2006)}


\title{Master equation for a quantum particle in a gas}

\begin{abstract}
  The equation for the quantum motion of a Brownian particle in a gaseous
  environment is derived by means of S-matrix theory. This quantum version of
  the linear Boltzmann equation accounts non-perturbatively for the quantum
  effects of the scattering dynamics and describes decoherence and dissipation
  in a unified framework. As a completely positive master equation it
  incorporates both the known equation for an infinitely massive Brownian
  particle and the classical linear Boltzmann equation as limiting cases.
\end{abstract}
\maketitle

How is the motion of a quantum particle affected by collisions with ambient
gas molecules? The well-established field of Quantum Brownian Motion
{\cite{qbmsum,Barnett2005a}} provides an answer provided the interaction can
be linearized and the particle state is close to classical. One is faced with
a rather different situation if the Brownian particle is in a highly
non-classical motional state, say, due to the passage through an
interferometer or the entanglement with another controlled degree of freedom.
Such correlations are becoming experimentally accessible in the emerging field
of molecular quantum optics, where the quantum nature of molecular motion is
tested and exploited {\cite{mqo}}.

In order to assess the (partial) loss of coherence in the case of strong,
nonclassical correlations in the motional state it is necessary to provide a
detailed, non-perturbative account of the microscopic scattering process. At
the same time, for the effects of decoherence to be relevant at all, the gas
density must be sufficiently low, so that the environmental gas can be safely
taken as not self-interacting and Markovian {\cite{Spohn1980a}}. In analogy to
the classical case {\cite{Cercignani1975a}} the associated description of a
{\tmem{single}} distinguished tracer particle within an ideal gas may be
called a {\tmem{linear}} Boltzmann equation. It should not be confused with
the linearized Boltzmann equations for the reduced single particle gas state,
obtained perturbatively from its multi-particle description.

The investigation of the loss of coherence due to gas collisions was initiated
by Joos and Zeh {\cite{Joos1985a}}, who considered the limiting case of an
infinitely massive tracer particle. This theory, which was later refined
{\cite{Gallis1990a,Hornberger2003b}} and tested experimentally
{\cite{colldecoexp}}, describes the pure spatial ``localization'' of an
extended coherent matter wave into a mixture with reduced spatial coherence,
but it cannot account for dissipation. The situation is much more involved if
the tracer mass $M$ is finite and comparable to the gas mass $m$ so that the
ratio $m / M$ must not be neglected. In this case ``localization'' occurs both
in position and in momentum \cite{Breuer2002a}, and the appropriate 
kinetic equation must
describe the full interplay of decohering and thermalizing dynamics.

So far, the most important advancement in this direction is the proposal by
Di\'osi {\cite{Diosi1995a}} of an equation based on a combination of
scattering theory and heuristic arguments. A more recent development is the
theory by Vacchini {\cite{VacchiniQBE}} in terms of the dynamic structure
factor of the medium, an approach limited to the (weak coupling) Born
approximation, like those in {\cite{Altenmuller1997a,Dodd2003a}}.

This letter presents the full quantum version of the linear Boltzmann
equation, describing the whole range of collisional effects from decoherence
to dissipation. It provides a transparent and stringent derivation, and a
discussion of its implications and limits. The only essential premise is the
Markov assumption, which implies{\emdash}in the spirit of Boltzmann's classic
derivation{\emdash}that both the rate and the effect of individual
two-particle scattering events are separately physically meaningful, while
subsequent collisions with the ``same'' gas molecule are negligibly unlikely.
Accordingly, the gas may be taken to be ideal (not self-interacting), 
stationary (diagonal
in momentum), and uniform in position space, thus covering
e.g. thermal 
states
of bosons and fermions, but no liquids.

In operator form the master equation reads $\partial_t \rho = \left( i \hbar
\right)^{- 1} \left[ \mathsf{H}, \rho \right] + \mathcal{L} \rho$ with
$\mathsf{H}=\mathsf{P}^2/(2M)$ the free Hamiltonian and
\begin{eqnarray}
  \mathcal{L} \rho & = & \bigintlim \mathd \tmmathbf{Q}
  \int_{\tmmathbf{Q}^{\bot}}  \frac{\mathd \tmmathbf{K}}{Q}  \left\{
  \mathsf{L} _{\tmmathbf{Q}, \tmmathbf{K}} \rho \mathsf{L} _{\tmmathbf{Q},
  \tmmathbf{K}}^{\dag}  \label{eq:qlbe} \right.\\
  &  & \left. - \frac{1}{2} \rho \mathsf{L} _{\tmmathbf{Q},
  \tmmathbf{K}}^{\dag} \mathsf{L} _{\tmmathbf{Q}, \tmmathbf{K}} - \frac{1}{2} 
  \mathsf{L} _{\tmmathbf{Q}, \tmmathbf{K}}^{\dag} \mathsf{L} _{\tmmathbf{Q},
  \tmmathbf{K}} \rho \right\} \bignone . \nonumber
\end{eqnarray}
Here the integration is over all momentum transfers $\tmmathbf{Q}$, and for
fixed $\tmmathbf{Q}$ also over the perpendicular plane $\tmmathbf{Q}^{\bot} =
\left\{ \tmmathbf{K} \in \mathbbm{R}^3 : \tmmathbf{K} \cdot \tmmathbf{Q}= 0
\right\}$. The Lindblad (jump) operators have the form
\begin{eqnarray}
  \mathsf{L} _{\tmmathbf{Q}, \tmmathbf{K}} & = & \mathe^{i \mathsf{R \cdot
  \tmmathbf{Q}/ \hbar}} F \left( \tmmathbf{K}, \mathsf{P} ; \tmmathbf{Q}
  \right)  \label{eq:Ldef}
\end{eqnarray}
where $\mathsf{R}$ and $\mathsf{P}$ are the position and the momentum operator
of the Brownian tracer particle. The function $F$, which is
operator-valued in (\ref{eq:Ldef}), contains all the details of
the collisional interaction with the gas. It involves the elastic scattering
amplitude $f \left( \tmmathbf{p}_{\tmop{out}}, \tmmathbf{p}_{\tmop{in}}
\right)$, the momentum distribution function $\mu \left( \tmmathbf{p} \right)$
of the gas{\footnote{E.g. for a Maxwell gas $\mu \left( \tmmathbf{p} \right) =
\exp \left( -\tmmathbf{p}^2 / p_T^2 \right) / \left( \pi^{3 / 2} p_T^3
\right)$ with $p^2_T = 2 mk_{\text{B}} T$ the most probable momentum.}}, and
its number density $n_{\tmop{gas}}$. It is convenient to denote relative
momenta by
\begin{eqnarray*}
  \tmop{rel} \left( \tmmathbf{p}, \tmmathbf{P} \right) & \assign &
  \frac{m_{\ast}}{m} \tmmathbf{p}- \frac{m_{\ast}}{M} \tmmathbf{P},
\end{eqnarray*}
with $m_{\ast} = mM / \left( M + m \right)$ the reduced mass. Moreover, for
given $\tmmathbf{Q} \neq 0$ let us denote the parallel and the perpendicular
contribution of a vector (operator) $\tmmathbf{P}$ by
$\tmmathbf{P}_{\|\tmmathbf{Q}} = \left( \tmmathbf{P} \cdot \tmmathbf{Q}
\right) \tmmathbf{Q}/ Q^2$ and by $\tmmathbf{P}_{\bot \tmmathbf{Q}}
=\tmmathbf{P}-\tmmathbf{P}_{\|\tmmathbf{Q}}$, respectively. With these
definitions
\begin{eqnarray}
  &  & F \left( \tmmathbf{K}, \tmmathbf{P}; \tmmathbf{Q} \right)
  \hspace{.5em} = \hspace{.5em} \frac{\sqrt{n_{\tmop{gas}} m}}{m_{\ast}}  
  \label{eq:Fdef}\\
  &  & \hspace{1em} \times f \left( \tmop{rel} \left( \tmmathbf{K}_{\bot
  \tmmathbf{Q}}, \tmmathbf{P}_{\bot \tmmathbf{Q}} \right) -
  \frac{\tmmathbf{Q}}{2}, \tmop{rel} \left( \tmmathbf{K}_{\bot \tmmathbf{Q}},
  \tmmathbf{P}_{\bot \tmmathbf{Q}} \right) + \frac{\tmmathbf{Q}}{2} \right)
  \nonumber\\
  &  & \hspace{1em} \times \mu \left( \tmmathbf{K}_{\bot \tmmathbf{Q}} +
  \left( 1 + \frac{m}{M} \right) \frac{\tmmathbf{Q}}{2} + \frac{m}{M}
  \tmmathbf{P}_{\|\tmmathbf{Q}} \right)^{1 / 2} . \nonumber
\end{eqnarray}
This implies that both the scattering amplitude and the distribution function
attain an operator character in (\ref{eq:Ldef}), and that the particle
momentum contributes with the part perpendicular to the momentum exchange to
the former and with the parallel one to the latter. For physically reasonable
interactions the function $F$ decreases sufficiently fast as $Q \rightarrow 0$
so that (\ref{eq:qlbe}) is well-defined.

Note that the form of $\mathcal{L} \rho$ fits the general structure of a
translation-invariant and completely positive master equation, as
characterized by Holevo {\cite{Holevo1996a}} (see {\cite{HolevoForm}} for a
discussion), although the summation in Ref.\,{\cite{Holevo1996a}} is here
replaced by the integrations in (\ref{eq:qlbe}). We will see below that the
master equation assumes a more intuitive form in the momentum representation.

My first aim is to provide a derivation of (\ref{eq:qlbe})--(\ref{eq:Fdef}).
To that end, let us first define the positive operator $\mathsf{\Gamma}$ which
yields the total collision rate. As in classical mechanics, the rate is
determined by the gas density, the modulus of the relative velocity $v =
\left| \tmop{rel} \left( \tmmathbf{p}, \tmmathbf{P} \right) \right| /
m_{\ast}$, and the total scattering cross section $\sigma \left(
\tmmathbf{p}_{\tmop{in}} \right)$. Denoting the improper momentum eigenvectors
of tracer and gas by $|\tmmathbf{P} \rangle$ and $|\tmmathbf{p} \rangle$ we
have
\begin{eqnarray}
  \mathsf{\Gamma} & = & \int_{} \mathd \tmmathbf{P} \mathd \tmmathbf{p}
  \bignone n_{\tmop{gas}} v \left( \tmmathbf{p}, \tmmathbf{P} \right) \sigma
  \left( \tmop{rel} \left( \tmmathbf{p}, \tmmathbf{P} \right) \right)
  |\tmmathbf{P} \rangle \langle \tmmathbf{P}| \otimes |\tmmathbf{p} \rangle
  \langle \tmmathbf{p}|. \nonumber
\end{eqnarray}
Indeed, for separable particle-gas states the expectation value of
$\mathsf{\Gamma}$ yields the average total collision rate experienced by the
Brownian particle.

Let us now see how a single collision changes the motional state of the tracer
particle according to scattering theory. If tracer and gas are uncorrelated
before the collision the outgoing tracer state is $\rho' =
\mathrm{\op{\tmop{tr}}}_{\tmop{gas}} \left( \mathsf{S} \left[ \rho \otimes
\rho_{\tmop{gas}} \right] \mathsf{S}^{\dagger} \right)$, where $\mathsf{S} =
\mathsf{I} + i \mathsf{T}$ is the two-particle scattering operator and a
partial trace over the gas has to be performed. Employing unitarity,
$\mathsf{S}^{\dagger} \mathsf{S} = \mathsf{I}$, one can express the change of
the state ${\Delta \rho = \rho' - \rho}$ as
\begin{eqnarray}
  \Delta \rho & = & \frac{i}{2} \tmop{tr}_{\tmop{gas}} \left( \left[
  \mathsf{T} + \mathsf{T}^{\dag}, \rho \otimes \rho_{\tmop{gas}} \right]
  \right)  \label{eq:Deltarho}\\
  &  & + \tmop{tr}_{\tmop{gas}} \left( \mathsf{T} \left[ \rho \otimes
  \rho_{\tmop{gas}} \right] \mathsf{T^{\dag}}  \right) \nonumber\\
  &  & - \frac{1}{2} \tmop{tr}_{\tmop{gas}} \left( \mathsf{T}^{\dag}
  \mathsf{T}  \left[ \rho \otimes \rho_{\tmop{gas}} \right] + \left[ \rho
  \otimes \rho_{\tmop{gas}} \right] \mathsf{T}^{\dag} \mathsf{T}  \right)
  \nonumber
\end{eqnarray}
The first term generates a constant coherent modification of the unitary
evolution and can be absorbed in the Hamiltonian $\mathsf{H}$. This energy
shift due to ``forward scattering'' is usually accounted for by a modified
index of refraction {\cite{Schmiedmayer1995short}} and can be
disregarded since the gas density is uniform.

The momentum representation of the remaining incoherent part of
(\ref{eq:Deltarho}) can be expressed entirely in terms of the kernel
\begin{eqnarray}
  &  & \langle \tmmathbf{P}| \tmop{tr}_{\tmop{gas}} \left( \mathsf{T} \left[
  |\tmmathbf{P}_0 \rangle \langle \tmmathbf{P}'_0 | \otimes \rho_{\tmop{gas}}
  \right] \mathsf{T^{\dag}} \right) |\tmmathbf{P}' \rangle .  \label{eq:Jdef}
\end{eqnarray}
If the momentum-diagonal representation of the gas state
\begin{eqnarray}
  \rho_{\tmop{gas}} & = & \frac{\left( 2 \mathpi \hbar \right)^3}{\Omega} \int
  \mathd \tmmathbf{p}_0 \mu \left( \tmmathbf{p}_0 \right) |\tmmathbf{p}_0
  \rangle \langle \tmmathbf{p}_0 | \bignone,  \label{eq:rhogas}
\end{eqnarray}
is inserted the kernel (\ref{eq:Jdef}) assumes the simple form $\delta \left(
\tmmathbf{P}-\tmmathbf{P}_0 -\tmmathbf{P}' +\tmmathbf{P}_0' \right) J \left(
\tmmathbf{P}, \tmmathbf{P}' ; \tmmathbf{P}-\tmmathbf{P}_0 \right)$, where the
function
\begin{eqnarray}
  J \left( \tmmathbf{P}, \tmmathbf{P}' ; \tmmathbf{Q} \right) & = &
  \frac{\left( 2 \pi \hbar \right)^3}{\Omega} \int \mathd \tmmathbf{p}_0 \mu
  \left( \tmmathbf{p}_0 \right) \bignone   \label{eq:Jtilde}\\
  &  & \times \langle \tmop{rel} \left( \tmmathbf{p}_0 -\tmmathbf{Q},
  \tmmathbf{P} \right) | \mathsf{T}_0 | \tmop{rel} \left( \tmmathbf{p}_0,
  \tmmathbf{P}-\tmmathbf{Q} \right) \rangle \nonumber\\
  &  & \times \langle \tmop{rel} \left( \tmmathbf{p}_0, \tmmathbf{P}'
  -\tmmathbf{Q} \right) | \mathsf{T}_0^{\dag} | \tmop{rel} \left(
  \tmmathbf{p}_0 -\tmmathbf{Q}, \tmmathbf{P}' \right) \rangle \nonumber
\end{eqnarray}
is given in terms of the single particle operator $\mathsf{T}_0$ for the
relative coordinates. Its momentum matrix elements are related to the
scattering amplitude by $\langle \tmmathbf{p}_f | \mathsf{T}_0 |
\tmmathbf{p}_i \rangle = \left( \pi \hbar \right)^{- 1} \delta \left(
\tmmathbf{p}_f^2 -\tmmathbf{p}_i^2 \right) f \left( \tmmathbf{p}_f,
\tmmathbf{p}_i \right)$ {\cite{Taylor1972a}}. However, inserting this into
(\ref{eq:Jtilde}) one arrives at an ill-defined expression, which involves the
arbitrary normalization volume $\Omega$ and, for for
$\tmmathbf{P}=\tmmathbf{P}'$, the {\tmem{square}} of a $\delta$-function. As
is well understood, the reason for this is our use of a momentum
representation of $\rho_{\tmop{gas}}$ in (\ref{eq:Jdef}). Plane waves are not
in the domain of the operator $\mathsf{S}_0 = \mathsf{I}_0 + i \mathsf{T}_0$,
which maps incoming asymptotes to outgoing ones (and by extension leaves the
outgoing ones invariant).

A possible resort would be therefore to choose a representation of
$\rho_{\tmop{gas}}$ which permits a decomposition into in- and out-states.
Indeed, a decomposition into Gaussian wave packets admits a conventional but
tedious calculation of the state change in the limiting case of an infinitely
massive Brownian particle, $m / M = 0$, as demonstrated in
{\cite{Hornberger2003b}}. In the same article it is shown that the identical
result can be obtained directly by keeping the diagonal representation. In
this case the extension of $\mathsf{T}_0$ beyond the physically acceptable
domain of incoming states must be complemented by a consistent, physically
motivated replacement rule:
\begin{eqnarray}
  \frac{\left( 2 \pi \hbar \right)^3}{\Omega} \left| \langle \tmmathbf{p}_f |
  \mathsf{T}_0 | \tmmathbf{p}_i \rangle \right|^2 & \rightarrow & \delta
  \left( \frac{\tmmathbf{p}_f^2 -\tmmathbf{p}_i^2}{2} \right)  \frac{\left| f
  \left( \tmmathbf{p}_f, \tmmathbf{p}_i \right) \right|^2}{\sigma \left(
  \tmmathbf{p}_i \right)  \left| \tmmathbf{p}_i \right|} 
  \label{eq:replacementrule}
\end{eqnarray}
In the present case of a finite mass ratio this yields a well defined kernel
(\ref{eq:Jtilde}) for $\tmmathbf{P}=\tmmathbf{P}'$, and thus implies $\int
\mathd \tmmathbf{Q} \tilde{J} \left( \tmmathbf{P}+\tmmathbf{Q},
\tmmathbf{P}+\tmmathbf{Q}; \tmmathbf{Q} \right) = 1$, which ensures the
conservation of the norm of $\rho'$. For $\tmmathbf{P} \neq \tmmathbf{P}'$ \
an extension of the rule (\ref{eq:replacementrule}) to different pairs of
incoming and outgoing relative momenta is required. It can be constructed at
no additional cost if the momentum change of the two pairs is the same, as is
the case in
\begin{eqnarray}
  X & = & \frac{\left( 2 \pi \hbar \right)^3}{\Omega} \langle \tmmathbf{p}_f
  +\tmmathbf{q}| \mathsf{T}_0 | \tmmathbf{p_i} +\tmmathbf{q} \rangle \langle
  \tmmathbf{p}_i -\tmmathbf{q}| \mathsf{T}_0^{\dag} |\tmmathbf{p}_f
  -\tmmathbf{q} \rangle g \left( \tmmathbf{q} \right) \nonumber
\end{eqnarray}
with arbitrary positive function $g$. Inserting the formal square root of the
replacement rule (\ref{eq:replacementrule}) yields the square root of a
product of two energy conserving $\delta$-functions with arguments $
\frac{\tmmathbf{p}_f^2 -\tmmathbf{p}_i^2}{2} \pm \left( \tmmathbf{p}_f
-\tmmathbf{p}_i \right) \cdot \tmmathbf{q}$. They imply that the parallel
component $\tmmathbf{q}_{\|} \equiv \tmmathbf{q}_{\| \left( \tmmathbf{p}_f
-\tmmathbf{p}_i \right)}$ of the momentum separation must be zero, which
restricts $\tmmathbf{q}$-integrations to the plane perpendicular to the
momentum change $\tmmathbf{p}_f -\tmmathbf{p}_i$. This restriction is equally
effected by replacing the vectors $\tmmathbf{q}$ with their projection
$\tmmathbf{q}_{\bot} \equiv \tmmathbf{q}-\tmmathbf{q}_{\|}$ onto that plane.
This way the form of the arguments in each individual scattering amplitude
already ensures the conservation of energy and one is left with a single
proper Dirac function $\delta \left( \frac{\tmmathbf{p}_f^2
-\tmmathbf{p}_i^2}{2} \right)$. Hence, as a natural generalization of
(\ref{eq:replacementrule}) we have
\begin{eqnarray}
  X & \rightarrow & \delta \left( \frac{\tmmathbf{p}_f^2 -\tmmathbf{p}_i^2}{2}
  \right)  \frac{f \left( \tmmathbf{p}_f +\tmmathbf{q}_{\bot}, \tmmathbf{p_i}
  +\tmmathbf{q}_{\bot} \right)}{\sqrt{\sigma \left( \tmmathbf{p_i}
  +\tmmathbf{q}_{\bot} \right)  \left| \tmmathbf{p_i} +\tmmathbf{q}_{\bot}
  \right|}}  \nonumber\\
  &  & \times \frac{f^{\ast} \left( \tmmathbf{p}_f -\tmmathbf{q}_{\bot},
  \tmmathbf{p_i} -\tmmathbf{q}_{\bot} \right)}{\sqrt{\sigma \left(
  \tmmathbf{p_i} -\tmmathbf{q}_{\bot} \right)  \left| \tmmathbf{p_i}
  -\tmmathbf{q}_{\bot} \right|}} g \left( \tmmathbf{q}_{\bot} \right) . 
  \label{eq:grr}
\end{eqnarray}
It turns into the known replacement rule as $\tmmathbf{q} \rightarrow 0$. The
only freedom in this construction is a possible phase factor from taking the
roots, but symmetry considerations lead to the above choice of no additional
phase.

Being able to evaluate traces with a momentum-diagonal representation of the
gas, we can now combine the operator $\mathsf{T}$ for the effect of a single
collision with the rate operator $\mathsf{\Gamma}$ to obtain the temporal
evolution (in interaction picture putting aside the contribution
of $\mathsf{H}$). The temporal change $\partial_t \rho$ is
obtained from $\Delta \rho$ in (\ref{eq:Deltarho}) by replacing $\mathsf{T}$ with $\mathsf{T}
\mathsf{\Gamma}^{1 / 2}$,
\begin{eqnarray}
  \partial_t \rho & \tilde{=} & \tmop{tr}_{\tmop{gas}} \left( \mathsf{T} 
  \mathsf{\Gamma}_{}^{1 / 2} \left[ \rho \otimes \rho_{\tmop{gas}} \right] 
  \mathsf{\Gamma}_{}^{1 / 2} \mathsf{T^{\dag}} \right) \nonumber\\
  &  & - \frac{1}{2} \tmop{tr}_{\tmop{gas}} \left(  \mathsf{\Gamma}_{}^{1 /
  2} \mathsf{T}^{\dag} \mathsf{T}  \mathsf{\Gamma}_{}^{1 / 2} \left[ \rho
  \otimes \rho_{\tmop{gas}} \right] \right) \nonumber\\
  &  & - \frac{1}{2} \tmop{tr}_{\tmop{gas}} \left(  \left[ \rho \otimes
  \rho_{\tmop{gas}} \right] \mathsf{\Gamma}_{}^{1 / 2} \mathsf{T}^{\dag}
  \mathsf{T}  \mathsf{\Gamma}_{}^{1 / 2} \right) .  \label{eq:Dtrho1}
\end{eqnarray}
This is suggested by a quantum trajectory unravelling of $\partial_t \rho$
{\cite{unrav}} where each trajectory is first weighted by the probability of a
collision event to take place in an infinitesimal time interval before being
scattered. To obtain the momentum representation of (\ref{eq:Dtrho1}) we have
to evaluate
\begin{equation}
  \langle \tmmathbf{P}| \tmop{tr}_{\tmop{gas}} \left(  \mathsf{T}
  \mathsf{\Gamma}_{}^{1 / 2}  \left[ |\tmmathbf{P}_0 \rangle \langle
  \tmmathbf{P}'_0 | \otimes \rho_{\tmop{gas}} \right] \mathsf{\Gamma}_{}^{1 /
  2} \mathsf{T^{\dag}} \right) |\tmmathbf{P}' \rangle \label{eq:Mdef}
\end{equation}
Using again the momentum diagonal gas state (\ref{eq:rhogas}) reduces the
kernel (\ref{eq:Mdef}) to the form $\delta \left( \tmmathbf{P}-\tmmathbf{P}_0
-\tmmathbf{P}' +\tmmathbf{P}_0' \right) M_{\tmop{in}} \left( \tmmathbf{P},
\tmmathbf{P}' ; \tmmathbf{P}-\tmmathbf{P}_0 \right)$. It follows that the
momentum representation of (\ref{eq:Dtrho1}) is
\begin{eqnarray}
  \partial_t \rho \left( \tmmathbf{P}, \tmmathbf{P}' \right) & = & \int \mathd
  \tmmathbf{Q} \bignone M_{\tmop{in}} \left( \tmmathbf{P}, \tmmathbf{P}' ;
  \tmmathbf{Q} \right) \rho \left( \tmmathbf{P}-\tmmathbf{Q}, \tmmathbf{P}'
  -\tmmathbf{Q} \right) \nonumber\\
  &  & - \frac{1}{2} \left[ M_{\tmop{out}}^{\tmop{cl}} \left( \tmmathbf{P}
  \right) + M_{\tmop{out}}^{\tmop{cl}} \left( \tmmathbf{P}' \right) \right]
  \rho \left( \tmmathbf{P}, \tmmathbf{P}' \right)   \label{eq:Dtrho2}
\end{eqnarray}
with
\begin{eqnarray}
  &  & M_{\tmop{in}} \left( \tmmathbf{P}, \tmmathbf{P}' ; \tmmathbf{Q}
  \right) \hspace{1em} = \hspace{1em} \int \mathd \tmmathbf{p}_0  \frac{\left(
  2 \mathpi \hbar \right)^3}{\Omega} \mu \left( \tmmathbf{p}_0 \right) 
  \frac{n_{\tmop{gas}}}{m_{\ast}}  \nonumber\\
  &  & \hspace{2em} \times \sqrt{\left| \tmmathbf{p}_i +\tmmathbf{q} \right|
  \sigma \left( \tmmathbf{p}_i +\tmmathbf{q} \right)}  \sqrt{\left|
  \tmmathbf{p}_i -\tmmathbf{q} \right| \sigma \left( \tmmathbf{p}_i
  -\tmmathbf{q} \right)} \bignone \nonumber\\
  &  & \hspace{2em} \times \langle \tmmathbf{p}_f +\tmmathbf{q}| \mathsf{T}_0
  |\tmmathbf{p}_i +\tmmathbf{q} \rangle \langle \tmmathbf{p}_i -\tmmathbf{q}|
  \mathsf{T}_0^{\dag} |\tmmathbf{p}_f -\tmmathbf{q} \rangle \nonumber
\end{eqnarray}
and $M_{\tmop{out}}^{\tmop{cl}} \left( \tmmathbf{P} \right) \assign \bigintlim
\mathd \tmmathbf{Q} \, M_{\tmop{in}} \left( \tmmathbf{P}+\tmmathbf{Q},
\tmmathbf{P}+\tmmathbf{Q}; \tmmathbf{Q} \right)$. Here I introduced
$\tmmathbf{p}_i \assign \tmop{rel} \left( \tmmathbf{p}_0,
\frac{\tmmathbf{P}+\tmmathbf{P}'}{2} -\tmmathbf{Q} \right)$ and
$\tmmathbf{p}_f \assign \tmmathbf{p}_i -\tmmathbf{Q}$ as functions of
$\tmmathbf{p}_0$, and defined {$\tmmathbf{q} \assign \tmop{rel} \left( 0,
\frac{\tmmathbf{P}-\tmmathbf{P}'}{2} \right)$}.

In order to evaluate $M_{\tmop{in}}$ the integration is now transformed to
$\mathd \tmmathbf{p}_i$. Incidentally, this suggests a natural factorization
of the $\mu$ distribution into a product of square roots, $\sqrt{\mu \left(
\tmmathbf{p}_0 \right) \mu \left( \tmmathbf{p}_0 \right)}$, since
$\tmmathbf{p}_0$ can be equally expressed as a function of $\tmmathbf{P}$ or
of $\tmmathbf{P}'$. Applying the replacement rule (\ref{eq:grr}) projects
$\tmmathbf{q}$ to $\tmmathbf{q}_{\bot} \equiv \tmmathbf{q}_{\bot \left(
\tmmathbf{p}_f -\tmmathbf{p}_i \right)}$ not only in the scattering
amplitudes, but also in the argument of $\mu$. One obtains the well-defined
expression \begin{widetext}
\begin{eqnarray}
  M_{\tmop{in}} \left( \tmmathbf{P}, \tmmathbf{P}' ; \tmmathbf{Q} \right) & =
  & \frac{n_{\tmop{gas}}}{m_{\ast}}  \left( \frac{m}{m_{\ast}} \right)^3 \int
  \mathd \tmmathbf{p}_i \delta \left( \frac{\tmmathbf{p}_f^2
  -\tmmathbf{p}_i^2}{2} \right) f \left( \tmmathbf{p}_f +\tmmathbf{q}_{\bot},
  \tmmathbf{p_i} +\tmmathbf{q}_{\bot} \right) f^{\ast} \left( \tmmathbf{p}_f
  -\tmmathbf{q}_{\bot}, \tmmathbf{p_i} -\tmmathbf{q}_{\bot} \right)
  \nonumber\\
  &  & \times \mu^{1 / 2} \left( \tmmathbf{p}_i + \frac{m}{M}  \left(
  \tmmathbf{p}_f +\tmmathbf{P} \right) + \frac{m}{m_{\ast}}
  \tmmathbf{q}_{\bot} \right) \mu^{1 / 2} \left( \tmmathbf{p}_i + \frac{m}{M} 
  \left( \tmmathbf{p}_f +\tmmathbf{P}' \right) - \frac{m}{m_{\ast}}
  \tmmathbf{q}_{\bot} \right) . \nonumber
\end{eqnarray}
\end{widetext} As the last step, the transformation $\tmmathbf{p}_i \rightarrow
\tmmathbf{K}= \frac{m}{m_{\ast}} \tmmathbf{p}_i + \frac{m}{M} 
\frac{\tmmathbf{P}_{\bot \tmmathbf{Q}} +\tmmathbf{P}_{\bot \tmmathbf{Q}}'}{2}
- \frac{m}{m_{\ast}}  \frac{\tmmathbf{Q}^{}}{2}$ factorizes the integrand into
$\tmmathbf{P}$ and $\tmmathbf{P}'$ contributions,
\begin{eqnarray}
  M_{\tmop{in}} \left( \tmmathbf{P}, \tmmathbf{P}' ; \tmmathbf{Q} \right) & =
  & \int \mathd \tmmathbf{K} \delta \left( \tmmathbf{K} \cdot \tmmathbf{Q}
  \right) F \left( \tmmathbf{K}, \tmmathbf{P}-\tmmathbf{Q}; \tmmathbf{Q}
  \right)  \nonumber\\
  &  & \phantom{\int} \times F^{\ast} \left( \tmmathbf{K}, \tmmathbf{P}'
  -\tmmathbf{Q}; \tmmathbf{Q} \right)  \label{eq:Min3}
\end{eqnarray}
with $F$ given in (\ref{eq:Fdef}). By returning to the Schr\"odinger
picture and noting $\int \mathd \tmmathbf{K} \delta \left(
  \tmmathbf{K} \cdot \tmmathbf{Q} \right) \bignone = Q^{- 1}
\int_{\tmmathbf{Q}^{\bot}} \mathd \tmmathbf{K}$ one finds that
(\ref{eq:Dtrho2}) with (\ref{eq:Min3}) is the momentum representation
of (\ref{eq:qlbe}), which closes its derivation.  Note that the form
of $\mathcal{L}$ is a consequence only of the premises and the
replacement (\ref{eq:replacementrule}).

The physics described by (\ref{eq:qlbe}) is easy to discuss in the momentum
basis (\ref{eq:Dtrho2}). We will see that $M_{\tmop{out}}^{\tmop{cl}} \left(
\tmmathbf{P} \right)$ is the rate of a classical particle with momentum
$\tmmathbf{P}$ to be scattered by the gas into a different direction or
velocity. Hence, the second line in (\ref{eq:Dtrho2}) effects a reduction of
the coherences {$\rho \left( \tmmathbf{P}, \tmmathbf{P}' \neq \tmmathbf{P}
\right)$} determined by the arithmetic mean of the corresponding momenta on
the diagonal. The first line in (\ref{eq:Dtrho2}), on the other hand, may
reduce or produce coherences. While the complex quantity $M_{\tmop{in}}$
cannot be decomposed into classical rates, it is reassuring that it involves
an integration over {\tmem{all}} possible, in general nonparallel, pairs of
two-particle scattering trajectories which end at the Brownian momenta
($\tmmathbf{P}, \tmmathbf{P}'$), each part conserving the total energy and the
momentum with an exchange of $\tmmathbf{Q}$, and weighted by the thus
restricted distribution of available gas momenta. [Di\'osi's equation
{\cite{Diosi1995a}}, in comparison, involves the differential cross section
$\mathd \sigma / \mathd \Omega = \left| f \right|^2$ and therefore at most
pairs of identical trajectories.]

After a time long compared to the time scale of decoherence the 
motional state is expected to be practically indistinguishable from a
classical state. As such it should be characterized by the momentum
distribution $w \left( \tmmathbf{P} \right) = \rho \left( \tmmathbf{P},
\tmmathbf{P} \right)$ alone, and one expects that the 
motion of the diagonal elements predicted by
(\ref{eq:qlbe}) is equal to the classical linear Boltzmann equation. Indeed,
one obtains from (\ref{eq:Dtrho2})
\begin{eqnarray}
  \partial_t w \left( \tmmathbf{P} \right) & = & \bigintlim \mathd
  \tmmathbf{Q}M^{\tmop{cl}}_{\tmop{in}} \left( \tmmathbf{P}, \tmmathbf{Q}
  \right) w \left( \tmmathbf{P}-\tmmathbf{Q} \right) \nonumber\\
  &  & \bignone - M^{\tmop{cl}}_{\tmop{out}} \left( \tmmathbf{P} \right) w
  \left( \tmmathbf{P} \right),  \label{eq:clbe}
\end{eqnarray}
where $M_{\tmop{out}}^{\tmop{cl}} \left( \tmmathbf{P} \right) = \bigintlim
\mathd \tmmathbf{Q}M_{\tmop{in}}^{\tmop{cl}} \left( \tmmathbf{P}+\tmmathbf{Q};
\tmmathbf{Q} \right)$ and
\begin{eqnarray}
  M^{\tmop{cl}}_{\tmop{in}} \left( \tmmathbf{P}; \tmmathbf{Q} \right) & =
  \bignone & \frac{n_{\tmop{gas}}}{m_{\ast}} \bigintlim \mathd \tmmathbf{K}
  \mu \left( \tmmathbf{K} \right) \delta \left( \frac{p_{cf}^2 - p_{ci}^2}{2}
  \right) \nonumber\\
  &  & \phantom{\frac{n}{m_{\ast}^2} \bigintlim} \times \frac{\mathd
  \sigma}{\mathd \Omega} \left( \tmmathbf{p}_{cf}, \tmmathbf{p}_{ci} \right) 
  \label{eq:Mclin}
\end{eqnarray}
is the rate density of a classical Brownian particle to be scattered into
momentum $\tmmathbf{P}$ upon a momentum exchange of $\tmmathbf{Q}$. Here
$\tmmathbf{p}_{ci} \equiv \tmop{rel} \left( \tmmathbf{K},
\tmmathbf{P}-\tmmathbf{Q} \right)$, $\tmmathbf{p}_{cf} \equiv
\tmmathbf{p}_{ci} -\tmmathbf{Q}$, and $\mathd \sigma / \mathd \Omega = \left|
f \right|^2$. Note that the classic form {\cite{Cercignani1975a}} of
(\ref{eq:clbe}) is obtained by transforming the $\tmmathbf{Q}$-integrals over
the Dirac $\delta$ into angular integrations over the relative momentum
direction. It follows that the stationary solution of (\ref{eq:qlbe}) is
given, for thermal gas states $\mu$, by the corresponding (momentum diagonal)
thermal Brownian state of the classical equation, and that the H-theorem
applies.

Another border case of (\ref{eq:qlbe}) is the limit of an infinitely massive
tracer. A short calculation confirms that letting $m / M$ approach zero
reduces $\mathcal{L} \rho$ to the (corrected version {\cite{Hornberger2003b}}
of the) master equation by Gallis and Fleming {\cite{Gallis1990a}}, which
attributes the loss of coherence to the amount of position information gained
by the colliding gas. Finally, (\ref{eq:qlbe}) assumes the form of Vacchini's
equation {\cite{VacchiniQBE}} if one replaces the true scattering amplitude
$f$ in (\ref{eq:Fdef}) by its Born approximation $f_B$ (which depends only on
the momentum transfer).

In conclusion, I presented the quantum version of the linear Boltzmann
equation. It unifies, in the form of a completely positive master equation,
the decohering and dissipative dynamics in the motion of a Brownian particle,
and it comprises various known dynamic behaviors as limiting forms.

I would like to thank John E.~Sipe and Bassano Vacchini for many helpful
discussions. This work was supported by the DFG Emmy Noether program.

\end{document}